\documentstyle[aps,epsfig]{revtex}
\begin{document}

\title{Folding model analysis of proton scattering from mirror nuclei $^{18}$Ne and $^{18}$O}

\author{D. Gupta\footnote{E-mail:dgupta@veccal.ernet.in}\footnote{Present Address: Institut de Physique Nucleaire, 91406 Orsay Cedex, France}, D. N. Basu}
\address{Variable  Energy  Cyclotron  Centre,  1/AF Bidhan Nagar,Kolkata 700 064, India}
\date{\today }
\maketitle

\begin{abstract}
The elastic and inelastic scattering of protons from mirror nuclei
$^{18}$Ne and $^{18}$O are studied in a folding model approach. For
comparison, two different effective interactions are folded with
Hartree-Fock densities to obtain the nuclear interaction potentials.
Both of them provide equivalent descriptions to the data and the
deformation parameters extracted from inelastic scattering are
reasonable. The density dependence parameters obtained from nuclear
matter calculations and used for present analysis also provide a
good estimate for the nuclear mean free path. The present formalism
unifies radioactivity, nuclear matter and nuclear scattering.

\vskip .2 true cm

\noindent
Keywords: Elastic and Inelastic Proton Scattering; Effective Interaction; Folding Model; SBM; DDM3Y; Mirror Nuclei

\noindent
PACS numbers: 25.40.Cm, 25.40.Ep, 21.30.Fe, 25.60.-t
\end{abstract}

\section {Introduction}
Proton scattering has been widely used as a means to study both collective and microscopic aspects of nuclear
structure~\cite{AM00,GU003}. The study is consistent only if a well-defined effective nucleon-nucleon (NN) interaction is applied
in the analysis. Also, with the advent of radioactive nuclear beams there is constant enhancement of our knowledge frontiers
on the structure and reaction dynamics of the known stable nuclei as well as their less known unstable
counterparts~\cite{TA85,GU002}. This rapidly developing field provides a testing ground for different nuclear reaction theories
and effective interactions. Scattering involving  $^{18}$Ne and $^{18}$O are interesting because not only are they mirror
nuclei, but also $^{18}$O is a stable nucleus while $^{18}$Ne is its radioactive counterpart.

In this work, proton scattering on $^{18}$Ne and $^{18}$O has been
studied at low energies ($<$100 MeV/A)~\cite{RI99,ES74} in a folding
model approach. The folding model is well known as a powerful tool
for analyzing nucleus-nucleus scattering data at relatively low
incident energies~\cite{GU002,PE93,RE74,SA97}. It directly links the
density profile of the nucleus with the scattering cross sections
and is thus very appropriate for studying nuclei, especially those
with exotic matter distributions. A semi-microscopic analysis in the
optical model (OM) framework is carried out. In the DWBA
calculations of nuclear excitation, with transferred angular
momentum $l$, the form factors are obtained by taking the
derivatives of the potentials used.

\section {Theoretical formulation}
The nucleon-nucleus potential can be obtained by single folding the density distribution of the
nucleus with the nucleon-nucleon  effective interaction~\cite{SA79} as,
\begin{equation}
U(\vec{r_1}) = \int \rho_2(\vec{r_2}) v_{\rm NN}(|\vec{r_1}-\vec{r_2}|)d^3\vec{r_2}\\
\end{equation}
where $\rho_2(\vec{r_2})$ is density of the nucleus at $\vec{r_2}$ and $v_{\rm NN}$ is the
effective interaction between two nucleons at the sites $\vec{r_1}$ and
$\vec{r_2}$. Two different forms of effective interactions have been employed in this work.
We perform a comparative study between the Modified Seyler-Blanchard (SBM)
and density dependent M3Y (DDM3Y) effective NN interactions.

The finite range, density, momentum and isospin dependent effective interaction SBM has
different strengths for pp (or nn) and pn interactions and its form is~\cite{BA90},
\begin{equation}
v(r = |\vec{r_1} - \vec{r_2}|,p,\rho) = -C_{l,u}\frac{e^{-r/a}}{r/a}[1 - \frac{p^2}{b^2} -d^2(\rho_1 +\rho_2)^n]
\end{equation}
where, the subscripts `$l$' and `$u$' refer to like-pair (nn or pp) and unlike-pair (np) interactions, respectively. Here `$a$' is
the range of the two-body interaction, `$b$' is a measure of the strength of repulsion with relative momentum `$p$', while
`$d$' and `$n$' are two parameters determining the strength of density dependence.
$\rho_1(\vec{r_1})$ and $\rho_2(\vec{r_2})$ are densities at the sites of the two interacting nucleons. The values of
parameters $n,~C_{\rm l},~C_{\rm u},~a,~b,~d$ are given in Table 1. These constants are found to reproduce the bulk
properties of nuclear matter and of finite nuclei~\cite{BA90,SA89} and are known also to explain the p +
$^{4,6,8}$He,$^{6,7,9,11}$Li scattering data successfully~\cite{GU003,GU002,KA95,KA97,KA98,KA99}. The parameters
are determined without exchange effects and thus they contain the effect indirectly though in a very approximate way.

The finite range M3Y effective interaction $v(r)$ appearing in Eqn.
1 is given by~\cite{r9}
\begin{equation}
 v(r) = 7999 \frac{e^{ - 4r}}{4r} - 2134 \frac{e^{ - 2.5r}}{2.5r}
\end{equation}
\noindent This interaction is based upon a realistic G-matrix, which
was constructed in an oscillator representation. Effectively it is
an average over a range of nuclear densities as well as energies and
therefore the M3Y has no explicit density or energy dependence. The
only energy dependent effect that arises from its use is a rather
weak one contained in an approximate treatment of single-nucleon
knock-on exchange. The density and energy averages are adequate for
the real part of the optical potential for heavy ions at lower
energies. Although, it is important to consider the density and
energy dependence explicitly for scattering at higher energies,
where the effects of a nuclear rainbow are seen and hence the
scattering becomes sensitive to the potential at small radii. Such
cases were studied introducing suitable and semirealistic explicit
density dependence \cite{r10,r11} into the M3Y interaction which was
then called the DDM3Y and was very successful for interpreting
consistently the high energy elastic $\alpha$ and heavy ion
scattering data. The present calculations use the density dependent
M3Y effective interaction (DDM3Y) supplemented by a zero-range
pseudo potential. In DDM3Y, the effective nucleon-nucleon
interaction $v(r)$ is assumed to be density and energy dependent and
therefore becomes functions of density and energy and is given by

\begin{equation}
  v(r,\rho,E) = t^{\rm M3Y}(r,E)g(\rho,E)
\end{equation}
\noindent where $t^{\rm M3Y}$ is the same M3Y interaction given by
Eqn. 3 but supplemented by a zero range pseudo-potential \cite{r10}

\begin{equation}
  t^{\rm M3Y} = 7999 \frac{e^{ - 4r}}{4r} - 2134\frac{e^{- 2.5r}}{2.5r} + J_{00}(E) \delta(r)
\end{equation}
\noindent
where the zero-range pseudo-potential representing the single-nucleon exchange term is given by

\begin{equation}
 J_{00}(E) = -276 (1 - 0.005E / A) {\rm MeV.fm^3}
\end{equation}
\noindent
and the density dependent part has been taken to be \cite{r11}

\begin{equation}
g(\rho, E) = C (1 - \beta(E)\rho^{2/3})
\end{equation}
\noindent
which takes care of the higher order exchange effects and the Pauli blocking effects. $E/A$ is energy per nucleon.
The constants of this interaction $C$ and $\beta$ when used in single folding model description, can be determined by nuclear matter calculations~\cite{BA04} as 2.07 and 1.624  fm$^2$ respectively.  The density dependence parameter $\beta$ has the
dimension of cross-section. The term $(1 - \beta\rho^{2/3})$ reduces the strength of the interaction and
changes sign at high densities making it repulsive. This is a direct consequence of the Pauli blocking effect. Thus
$(1 - \beta\rho^{2/3})$ represents the probability of non-interaction arising due to collision probability $\beta\rho^{2/3}$ of a
nucleon in nuclear medium of density $\rho$. The parameter $\beta$ can be identified as the `in medium'
effective nucleon-nucleon interaction cross section $\sigma_0$. This value of  $\beta$ along with nucleonic density of infinite nuclear
matter $\rho_0$ can also provide the nuclear mean free path $\lambda=(\rho_0 \sigma_0)^{-1}$.

\section {Calculation and Analysis}

The nuclear ground state densities have been calculated in the framework of spherical Hartree Fock plus BCS
calculations in co-ordinate space using two different parameter sets of the Skyrme interactions given by~\cite{BA82}
 \begin{equation}
v_{\rm central}({\bf r_1,r_2}) = t_0(1 + x_0{\bf P}_{\sigma})\delta({\bf r}) + t_1(1 + x_1{\bf P}_{\sigma})[\delta({\bf r}) {\bf k^2} +
{\bf {k'}^2}\delta({\bf r})] + t_2(1 + x_2{\bf P}_{\sigma}) {\bf k'}\delta({\bf r}) {\bf k} + t_3(1 + x_3{\bf P}_{\sigma})\rho^{\alpha}
({\bf R})\delta({\bf r})
\end{equation}
There are negligible differences in the ground state densities of
the nuclei with the two different parameter sets (for
SIII~\cite{BE75} and SkM*~\cite{BA82}) provided in Table 2. We show
SkM* parameterization for the $^{18}$O ground state density (Fig. 1)
and used it for calculations of the single folding potential and
form factor. We carried out the same procedure for $^{18}$Ne and for
SkM* the binding energies per nucleon are in good agreement (within
0.1 MeV) with the experimental values for  both the nuclei.

The potentials used while calculating phenomenological best
fits~\cite{ES74,AL99} have the following form,
\begin{equation}
V_{\rm pheno}(r) = -V_o~f_o(r)~-~i~W_v~f_v(r) + 4~i~a_sW_s(d/dr)
f_s(r) + 2(\hbar/m_{\pi}c)^2~V_{s.o} 1/r (d/dr) f_{s.o}(r)~({\bf
L.S}) + V_{\rm coul}
\end{equation}
where, $f_x(r)~= [1~+~exp(\frac{r-R_x}{a_x})]^{-1}$,
$R_{x}~=~r_{x}A^{1/3}$ and $x~=~o,v,s,s.o$. The subscripts
$o,v,s,s.o$ denote real, volume imaginary, surface imaginary and
spin-orbit respectively and  $V_o$, $W_v$ ($W_s$) and $V_{s.o}$ are
the strengths of the real, volume (surface) imaginary and spin-orbit
potentials respectively. $V_{\rm coul}$ is the Coulomb potential of
a uniformly charged sphere of radius 1.20~$A^{1/3}$.

In semi-microscopic analysis both the volume real ($V$) and volume
imaginary ($W$) parts of the potentials (generated microscopically
by folding model) are assumed to have the same shape, as in
Ref.~\cite{GU002}, i.e. $V_{\rm micro}(r)~=~V + iW$ = ($N_{\rm R}$ +
$iN_{\rm I}$)$U$($r_1$) where, $N_{\rm R}$ and $N_{\rm I}$ are the
renormalization factors for real and imaginary parts
respectively~\cite{SA97}. In Fig. 2 the p + $^{18}$O renormalized
real folded  potentials (employing the SBM and DDM3Y
interactions) are shown at $E$ = 24.5$A$ MeV, along with the best
fit phenomenological real part. Thus the  potentials for elastic
scattering analysis include real and volume imaginary terms (folded
potentials) and also surface imaginary and spin-orbit terms (best
fit phenomenological potentials).

For each angular distribution, best fits are obtained by minimizing
$\chi^2$/N, where $\chi^2$ = $\sum_{k = 1}^{\rm N}
\left[\frac{\sigma_{th}(\theta_k)~-\sigma_{ex}(\theta_k)}{\Delta\sigma_{ex}
(\theta_k)}\right]^2$, where $\sigma_{th}$ and $\sigma_{ex}$ are the
theoretical and experimental cross sections respectively, at angle
$\theta_k$, $\Delta\sigma_{ex}$ is the experimental error and N is
the number of data points. The potentials for elastic scattering
analysis are subsequently used in the DWBA calculations of inelastic
scattering with transferred angular momentum $l$. The calculations
are performed using the code DWUCK4~\cite{DWUCK4}. The derivative of
the potentials ($\delta \frac{dV}{dr}$) are used as the form
factors. The microscopic real and imaginary form factors have the
same shape with strengths $N_{\rm R}^{\rm FF}$ and $N_{\rm I}^{\rm
FF}$ respectively, where $N_{\rm R,I}^{\rm FF}$ = $N_{\rm R,I}r_{\rm
rms}^V$, where the radius parameter $r_{\rm rms}^V$ is the rms
radius of the folded potential. In addition, form factors derived
from phenomenological surface imaginary and spin-orbit potentials
are included. The deformation parameters $\delta$ are determined by
fitting the inelastic scattering angular distribution. The
renormalizations required for the potentials are reminiscent of
those for deuteron and $^6$Li scattering and it may be surmised that
it is  for the same reasons; weak binding and ease of breakup and
other reaction channels. Table 1 gives the parameters  of the
interactions  used here. Both interactions provide incompressibility
of $\sim$ 300 MeV for spin and isospin symmetric cold infinite
nuclear matter. Moreover, in case of DDM3Y, $\beta$ value of 1.624
fm$^2$ obtained from nuclear matter calculations, along with
nucleonic density of 0.16 fm$^{-3}$ provides a mean free path of
about 4 fm which is in excellent agreement with other theoretical
estimates~\cite{SI83}. Table 3 gives the phenomenological best fit
optical model parameters while Table 4 gives the renormalization
factors, $\delta$ values, $\chi^2$/N  for the folding model
analysis.

The relationship between the reduced electric quadrupole transition
rate $B(E2)$ for the ground state to the $2^+$ state in units of $e^2fm^4$ and
the quadrupole deformation parameter $\delta$ is given by \cite{RA87}

\begin{equation}
\delta(1+0.16\delta+0.20\delta^2+....)= 4 \pi B(E2)^{1/2}/(3ZR^2)
\end{equation}
\noindent where $R=1.2A^{1/3}$ fm and  $Z$ is the atomic number. The
quadrupole deformations listed in reference~\cite{RA87} were
obtained by using Eqn. 10 but keeping only the terms up to first
order in $\delta$. We have recalculated these values by keeping
terms up to third order in $\delta$. The recalculated quadrupole
deformations thus obtained from the experimental $B(E2)$ values
listed in reference~\cite{RA87} are 0.33078 and 0.59284 for $^{18}$O
and $^{18}$Ne respectively. As can be seen from Table 4, the
quadrupole deformation obtained from the present analysis for
$^{18}$O is in excellent agreement while that for $^{18}$Ne is
significantly underestimated due to lack of experimental data at
forward angles. The inelastic scattering is more sensitive at
forward angles due to its relative purity compared to data
corresponding to backward angular range where other non-elastic
processes also contribute. The quality of elastic as well as the
inelastic fits deteriorate at backward angles and similar
deterioration of fits are also seen for proton scattering from other
nuclei~\cite{SA79,KH02}. The reason for this is probably that the
full cross section is ascribed to potential scattering while
quasi-compound-nucleus formation feeds back into the elastic channel
and whose energy-dependence is controlled by barrier-top resonances.
The backward angular range classically corresponds to smaller impact
parameters. This fact suggests higher compound nuclear formation
probabilities at backward angles while those at forward angles are
expected to be negligibly small. Since relative contributions of
compound elastic and direct elastic are not disentangled, $\chi^2$
was calculated only upto about 90$^o$ in center of mass during
fitting the $^{18}$O data.

It may be noted that since we could not acquire the experimental data
for $^{18}$O, they have been read quite accurately from the plots in the original paper~\cite{ES74}
and a 5$\%$ uniform error has been assumed. The technique used for extracting the data from plots
along with relevant co-ordinate transformations are described in the appendix.

\section {Conclusion}
In the present study we find that the parametrized SBM effective interaction, and the realistic DDM3Y
effective interaction obtained from sophisticated G-matrix calculations provide equivalent
descriptions for the elastic and inelastic scattering of protons from the mirror nuclei $^{18}$Ne and $^{18}$O
(Fig. 3, 4). The values of the deformation parameters have been extracted from the calculations for
these nuclei. Even though the analysis reported here is quite detailed,
measurements at higher energies may still be useful in distinguishing various effective interactions.
The lack of sensitivity is due to the fact that at such energies, nuclear densities primarily probed are near the
surface, which makes the density dependent effects less realizable. The
form of density dependence used here is more physical as compared to other forms.
The parameter $\beta$ can be interpreted  as `in medium'
nucleon-nucleon  interaction cross section while  $(1 - \beta\rho^{2/3})$  as the non-interaction
probability  arising due to higher order exchange and Pauli blocking effects.

In summary,  a consistent folding model analysis of proton
scattering on A = 18 nuclei is carried out using two different
effective NN interactions. The conventional way of generating the
form factors is followed, that is, by taking the derivatives of the
potentials (microscopic real and imaginary as well as
phenomenological surface imaginary and spin-orbit potentials).
Deformation parameters ($\delta$) are extracted from the analyses.
The results obtained for the deformation parameter are in good
agreement with the available results. The density dependence
parameter obtained from nuclear matter calculations, which has been
used in the single folding model description for the analysis of
elastic and inelastic scattering of protons, also provides excellent
estimate for nuclear mean free path. We know that the DDM3Y
effective interaction, has profound theoretical basis. It provides
unified description of cluster radioactivity, scatterings of
$\alpha$ and heavy ion when used in a double folding model, and
nuclear matter when used in a single folding model. We find that it
also provides reasonable description for elastic and inelastic
scattering of protons.

The authors gratefully acknowledge L. A. Riley for sending the experimental data in a tabular form.

\begin{table}
{\bf Table 1:}\\
Parameters of the SBM and DDM3Y interactions\\
\setlength{\tabcolsep}{5.0 mm}
\begin{tabular}{ccccccc}
\hline\hline \multicolumn{1}{c}{SBM}& \multicolumn{1}{c}{$n$}&
\multicolumn{1}{c}{$C_l$(MeV)}& \multicolumn{1}{c}{$C_u$(MeV)}&
\multicolumn{1}{c}{$a$(fm)}& \multicolumn{1}{c}{$b$(MeV/c)}&
\multicolumn{1}{c}{$d$(fm)}\\
\hline
&2/3&215.7&669.3&0.554&668.7&0.813\\
\hline
DDM3Y&$n$&$C$&$\alpha$(MeV$^{-1})$&$\beta$(fm$^2$)&&\\
\hline &2/3&2.07&0.005&1.624&&\\ \hline\hline
\end{tabular}
\end{table}

\begin{table}
{\bf Table 2:} \\
Parameter sets of the Skyrme interactions\\
\setlength{\tabcolsep}{3.0 mm}
\begin{tabular}{c|c|c|c|c|c|c|c|c|c|c}
\hline \hline \multicolumn{1}{c}{Interaction}&
\multicolumn{1}{c}{$t_0$}& \multicolumn{1}{c}{$t_1$}&
\multicolumn{1}{c}{$t_2$}& \multicolumn{1}{c}{$t_3$}&
\multicolumn{1}{c}{$t_4$}& \multicolumn{1}{c}{$x_0$}&
\multicolumn{1}{c}{$x_1$}& \multicolumn{1}{c}{$x_2$}&
\multicolumn{1}{c}{$x_3$}&
\multicolumn{1}{c}{$\alpha$}\\
\hline
SIII&-1128.75&395.0&-95.0&14000.0&120.0&0.45&0.0&0.0&1.0&1\\
SkM*&-2645.0&410.0&-135.0&15595.0&130.0&0.09&0.0&0.0&0.0&1/6\\
\hline \hline
\end{tabular}
\end{table}

\begin{table}
{\bf Table 3:} \\
Phenomenological potential parameters used in p +  $^{18}$Ne and $^{18}$O scattering \\
\setlength{\tabcolsep}{0.5 mm}
\begin{tabular}{c|c|c|c|c|c|c|c|c|c|c|c|c|c|c|c|l}
\hline \hline \multicolumn{1}{c|}{Nucleus}&
\multicolumn{1}{c|}{$E/A$}& \multicolumn{1}{c}{$V_o$}&
\multicolumn{1}{c}{$r_o$}& \multicolumn{1}{c|}{$a_o$}&
\multicolumn{1}{c}{$W_v$}& \multicolumn{1}{c}{$r_v$}&
\multicolumn{1}{c|}{$a_v$}& \multicolumn{1}{c}{$W_s$}&
\multicolumn{1}{c}{$r_s$}& \multicolumn{1}{c|}{$a_s$}&
\multicolumn{1}{c}{$V_{s.o}$}& \multicolumn{1}{c}{$r_{s.o}$}&
\multicolumn{1}{c|}{$a_{s.o}$}& \multicolumn{1}{c|}{$\chi^2_{\rm
el}$/N}& \multicolumn{1}{c|}{$J/A$}&
\multicolumn{1}{l}{ Ref.}\\
&(MeV)&(MeV)&(fm)&(fm)&(MeV)&(fm)&(fm)&(MeV)&(fm)&(fm)&(MeV)&(fm)&(fm)&&(MeV fm$^3$)&\\
\hline
$^{18}$Ne&30.0&40.00&1.100&0.730&&&&7.00&1.380&0.600&7.80&1.090&0.740&5.265&-364.7&\cite{ES74}\\
$^{18}$O&24.5&48.57&1.163&0.780&2.266&1.169&0.690&5.77&1.169&0.690&5.90&0.882&0.630&6.896&-527.0&\cite{AL99}\\
\hline\hline
\end{tabular}
\end{table}

\begin{table}
{\bf Table 4:} \\
Renormalizations of
SBM and DDM3Y folded potentials and form factors for p +  $^{18}$Ne and $^{18}$O scattering at incident energy ($E/A$) and
excited state energy ($E^*$) in MeV, angular momentum transfer ($l$), deformation parameter ($\delta$), volume integral
($J/A$) of the real folded potential in MeV fm$^3$ and $\chi^2$/N values from
best-fits to the elastic and inelastic scattering data\\
\setlength{\tabcolsep}{2.0 mm}
\begin{tabular}{cccccccccccccc}
\hline \hline \multicolumn{1}{c}{Nucleus}&
\multicolumn{1}{c}{$E/A$}& \multicolumn{1}{c}{$E^*$}&
\multicolumn{1}{c}{$N_{\rm R}$}& \multicolumn{1}{c}{$N_{\rm I}$}&
\multicolumn{1}{c}{$r_{\rm rms}^V$}& \multicolumn{1}{c}{$N_{\rm
R}^{\rm FF}$}& \multicolumn{1}{c}{$N_{\rm I}^{\rm FF}$}&
\multicolumn{1}{r}{$l$}& \multicolumn{1}{c}{$\delta$}&
\multicolumn{1}{r}{$\chi^2_{\rm el}$/N}&
\multicolumn{1}{r}{$\chi^2_{\rm inel}$/N}&
\multicolumn{1}{r}{$J/A$}&
\multicolumn{1}{r}{Interaction}\\
\hline
$^{18}$Ne$^*$&30.0&1.890&0.67&0.000&3.462&2.320&0.000&2&0.436&5.992&0.973&-336.1&SBM\\
$^{18}$Ne$^*$&30.0&1.890&0.62&0.000&3.538&2.194&0.000&2&0.400&5.798&0.859&-358.0&DDM3Y\\
\hline
$^{18}$O$^*$&24.5&1.982&0.87&0.006&3.547&3.086&0.020&2&0.400&7.131&4.471&-502.7&SBM\\
$^{18}$O$^*$&24.5&1.982&0.80&0.006&3.508&2.806&0.019&2&0.367&6.704&7.696&-469.2&DDM3Y\\
\hline\hline
\end{tabular}
\end{table}

\newpage
\section{Appendix}

This appendix is aimed at providing a simple co-ordinate transformation
formula for general purpose use in extracting data from plots using computer graphics.

In many instances the values of the experimental data are not available in
literatures while the plots are shown. Particularly, for old plots, tabulated
data are often very difficult to acquire. In such cases, data values can be
read in quite accurately by using computer programs. In some cases
both linear and logarithmic scales are involved. A general formula
is thus required which will convert the co-ordinates of the data points
as shown by the software to their actual values. It should also
be taken into account that the co-ordinate system of the printed
graph may be rotated, albeit small, with respect to the software
co-ordinate system. Thus a handy formula would be extremely useful
for such purposes.

We assume ($X,Y$) to be the co-ordinates of the printed graph while ($x,y$) are
those of the computer program such as GSview. In Fig. 5, ($x_0,y_0$) is the origin of
the printed graph while ($x_1,y_1$) and ($x_2,y_2$) are two points on the $X$ and
$Y$ axes, respectively.

Since the ratio of two linear lengths measured in one co-ordinate system should be equal
to that in the other system,
{\Large $\frac{X-X_0}{X_1-X_0} =  \frac{(x - x_0)sec\theta + [(y - y_0) - (x - x_0)tan\theta]sin\theta}{(x_1 - x_0)sec\theta}$} 
where, $tan\theta$ ={\Large $\frac{y_1-y_0}{x_1-x_0}$}. A little manipulation gives,
\begin{equation}
X  = X_0 + (X_1-X_0)\Bigl[\frac{(x - x_0)cos\theta + (y - y_0)sin\theta}{(x_1 - x_0)sec\theta}\Bigr]
\end{equation}

We consider X scale to be linear and Y scale to be logarithmic. Therefore,\\
{\Large $\frac{logY-logY_0}{logY_2-logY_0} =  \frac{[(y - y_0) - (x - x_0)tan\theta]cos\theta}{(y_2 - y_0)sec\theta}$}
which gives,
\begin{equation}
Y  = Y_0\Bigl(\frac{Y_2}{Y_0}\Bigr)^{\frac{-(x - x_0)sin\theta + (y - y_0)cos\theta}{(y_2 - y_0)sec\theta}}
\end{equation}
If both axes are linear, then Eqn. 11 holds while
\begin{equation}
Y  = Y_0 + (Y_2-Y_0)\Bigl[\frac{-(x - x_0)sin\theta + (y - y_0)cos\theta}{(y_2 - y_0)sec\theta}\Bigr]
\end{equation}
If both axes are logarithmic then Eqn. 12 holds while,
\begin{equation}
X  = X_0\Bigl(\frac{X_1}{X_0}\Bigr)^{\frac{(x - x_0)cos\theta + (y - y_0)sin\theta}{(x_1 - x_0)sec\theta}}
\end{equation}
The expressions provided above, though simple, will be immensely useful for experimentalists as well as
theoreticians for extracting data where their explicit values are not available.

\begin{figure}[h]
\eject\centerline{\epsfig{file=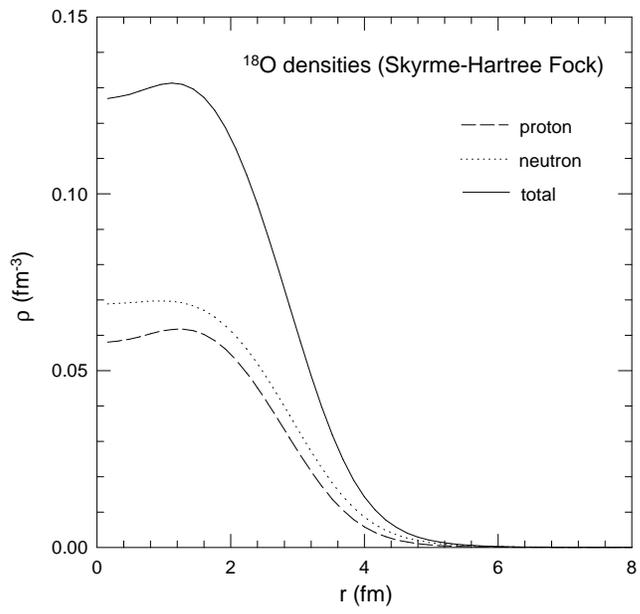,height=20cm,width=14cm}}
\caption{Skyrme Hartree-Fock densities of $^{18}$O used in this
work.}
\label{fig1}
\end{figure}

\begin{figure}[h]
\eject\centerline{\epsfig{file=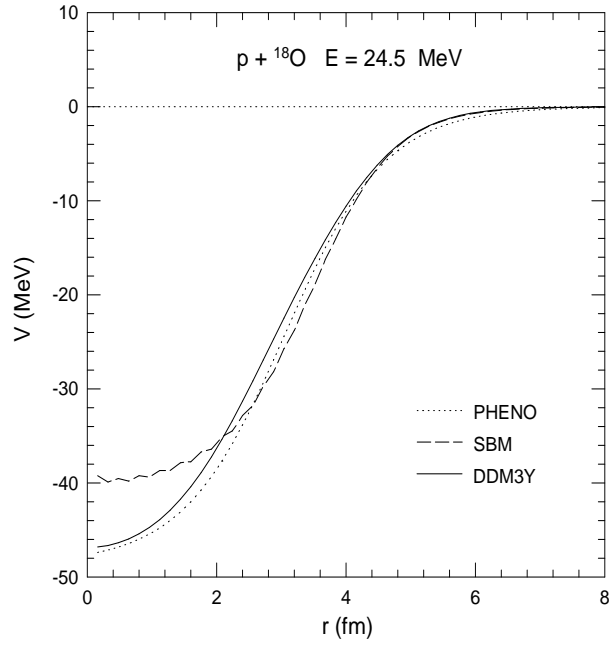,height=20cm,width=14cm}}
\caption{The phenomenological and renormalized real folded potentials (employing SBM and DDM3Y interactions)
for p + $^{18}$O at 24.5A MeV.}
\label{fig2}
\end{figure}

\begin{figure}[h]
\eject\centerline{\epsfig{file=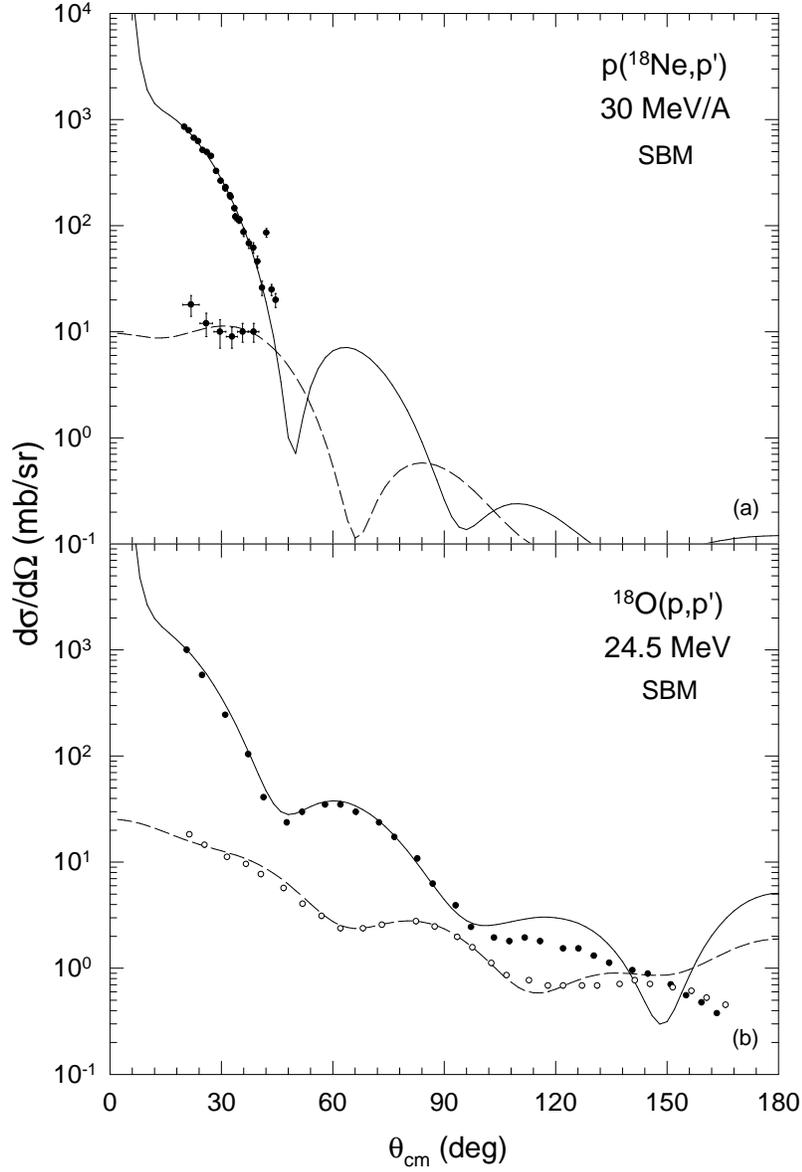,height=20cm,width=14cm}}
\caption {The experimental angular distributions and folding model calculations (SBM) of (a) p + $^{18}$Ne at 30A MeV for elastic and inelastic [$E^*$ = 1.890 MeV (2$^+$)] scattering [5], (b) p + $^{18}$O at 24.5A MeV for elastic and inelastic [$E^*$ = 1.982 MeV (2$^+$)] scattering [6]. The corresponding $N_{\rm R}$, $N_{\rm I}$, $N_{\rm R}^{\rm FF}$, $N_{\rm I}^{\rm FF}$ values and phenomenological surface imaginary and spin-orbit parameters are given in Tables 3, 4. The continuous and  dashed lines correspond to calculations for elastic and inelastic cross sections respectively.}
\label{fig3}
\end{figure}

\begin{figure}[h]
\eject\centerline{\epsfig{file=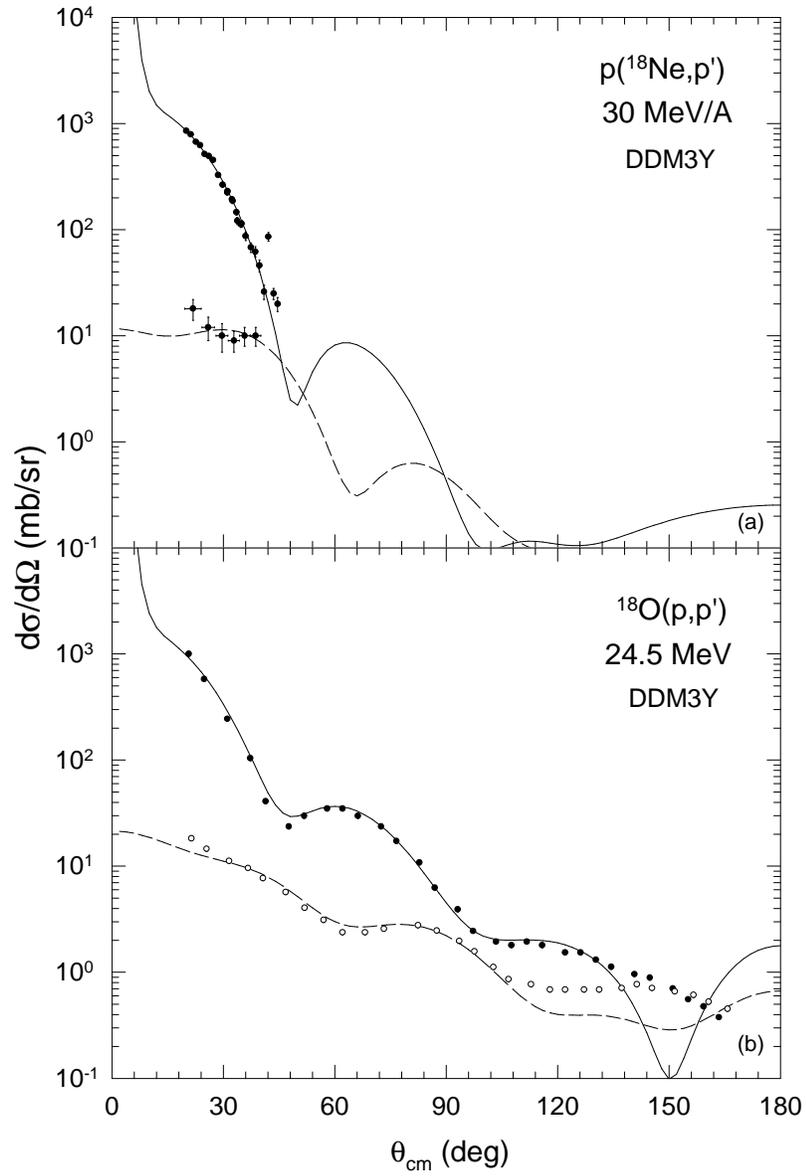,height=20cm,width=14cm}}
\caption{The same as in Fig. 3 but employing DDM3Y interaction.}
\label{fig4}
\end{figure}

\begin{figure}[h]
\eject\centerline{\epsfig{file=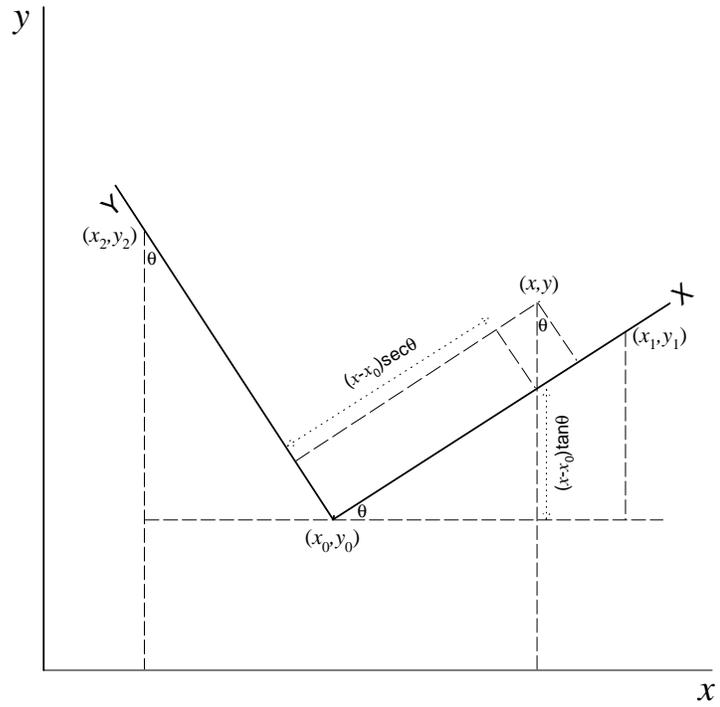,height=20cm,width=14cm}}
\caption {($X,Y$) are the printed graph co-ordinate system, rotated
by an angle $\theta$ with respect to ($x,y$), the computer program
co-ordinate system.}
\label{fig5}
\end{figure}

\end{document}